\documentclass[runningheads]{llncs}

\usepackage{header}	

\usepackage{symbols}	

\title{What Programs Want}
\subtitle{Automatic Inference of Input Data 
Specifications}
\author{Caterina Urban\inst{1, 2}}
\institute{INRIA, Paris, France
	\and DIENS, \'Ecole Normale 
	Sup\'erieure, CNRS, PSL University, Paris, France
	\\
	\email{caterina.urban@inria.fr}}
\date{}

\begin{document}

\maketitle

\begin{abstract}
	Nowadays, as machine-learned software quickly
	permeates our society, we are becoming 
	increasingly vulnerable to programming errors in 
	the data pre-processing or training software, as 
	well as errors in the data itself. 
	In this paper, we propose a static shape analysis 
	framework for input data of data-processing 
	programs. Our analysis 
	automatically infers 
	necessary conditions on the structure and values 
	of the data read by a data-processing program. 
	Our framework builds on a family of 
	underlying abstract domains, extended to 
	indirectly reason about the input data rather than 
	simply reasoning about the program variables. 
	The choice of these abstract domain is a 
	parameter of the analysis. 
	We describe various instances built from existing 
	abstract domains. 
	The proposed approach is implemented in an 
	open-source static analyzer for \python 
	programs. We demonstrate its potential
	on a number of representative examples.
\end{abstract}

\section{Introduction}

Due to the availability of vast amounts 
of data and corresponding tremendous advances in 
machine learning, computer software is nowadays 
an ever 
increasing presence in every aspect 
our society.
As we rely more and more on machine-learned 
software, we become increasingly vulnerable to 
programming errors but (in contrast to traditional 
software) also errors in the data used 
for training.

In general, before software training, the data goes 
through 
long pre-processing
pipelines\footnote{\url{https://www.nytimes.com/2014/08/18/technology/for-big-data-scientists-hurdle-to-insights-is-janitor-work.html}}.
Errors can be 
missed, or even introduced, at any stage of these 
pipelines.
This is even more true when data pre-processing 
stages are disregarded as single-use glue code and, 
for this reason, are poorly tested, let alone statically 
analyzed or verified. 
Moreover, this kind of
code is 
often written in a rush and is highly dependent on 
the data (e.g., the use of magic constants is not 
uncommon)
All this together, greatly increases 
the likelihood for errors to be noticed extremely 
late in the pipeline (which entails a more or less 
important waste of 
time), or more dangerously, to remain completely 
unnoticed.

\subsubsection{Motivating Example.}


\begin{figure}[t]
\begin{lstlisting}[language=Python]
grade$2$gpa = { 'A': 4.0, 'B': 3.0, 'C': 2.0, 'D': 1.0, 'F': 0.0 }
students = int(input())
for _ in range(students):
    name = input()
    classes = int(input())
    gpa = 0.0
    for _ in range(classes):
        grade = input()
        gpa += grade$2$gpa[grade]
    result = gpa / classes
    print('{}: {}'$.$format(name, result))
\end{lstlisting}
	\caption{Simple GPA calculator for multiple 
		students.}
	\label{fig:gpa}
\end{figure}

As an example, let us consider the data 
processing \python code shown in 
Figure~\ref{fig:gpa}, which
calculates the simple GPA for a 
given number of students (cf. Line $2$). 
For each class taken by a student (cf. Line $7$), 
their (A-F) grade is converted into a numeric (4-0) 
grade, and all numeric grades are added together 
(cf. Line $9$). 
The GPA is obtained by 
dividing this by the number of 
classes taken by the student (cf. Line $10$).

Even this small program makes several 
assumptions on its input data. For instance, it 
assumes that the very first input read by the 
program (cf. Line $2$) is a string representation of 
an integer number that indicates how many student 
records follow in the data file (cf. Line $3$). 
A similar assumption holds for the second input 
read for each student 
record (cf. Line $5$), which should indicate how 
many 
student grades follow in the data file (cf. Line $7$). 
This number should be different from zero (or the 
division at Line $10$ would raise a 
\lstinline[language=Python]{ZeroDivisionError}). 
Finally, the program assumes that each grade read 
at Line $8$ is a 
string in the set 
$\set{\text{\lstinline[language=Python]{'A'}}, 
	\text{\lstinline[language=Python]{'B'}}, 
	\text{\lstinline[language=Python]{'C'}}, 
	\text{\lstinline[language=Python]{'D'}}, 
	\text{\lstinline[language=Python]{'F'}}}$ (or the 
dictionary access at Line 
$9$ would raise a 
\lstinline[language=Python]{KeyError}). Note that, 
not all assumptions necessarily lead to a program 
error if 
violated. 
For 
instance, consider the following data stream:
\begin{center}
	\begin{tabular}{ccccc}
		1 & Emma & 1 & A & F \\
		&& $\shortuparrow$ && 
	\end{tabular}
\end{center}
A mistake is indicated by the arrow: the 
number of classes taken by the student Emma is off 
by one (i.e., it should be $2$ instead of $1$). In this 
case 
the program in Figure~\ref{fig:gpa} will not raise 
any error but will instead compute a wrong (but 
plausible!) GPA for 
Emma (i.e., $4.0$ instead of $2.0$).

\subsubsection{Our Approach.}

\looseness=-1
To address these issues, we propose 
an abstract interpretation-based \emph{shape 
analysis} framework \emph{for 
input data} of 
data-processing programs. The analysis 
automatically 
infers implicit assumptions on the 
input data that are embedded in the source 
code of a program. 
Specifically, we infer assumptions on the 
structure of the data as well as on the 
values 
and the relations between the data. 

We propose a \emph{new} data shape 
\emph{abstract domain}, capable of reasoning 
about the 
input data in addition to the program variables. The 
domain builds on a family of underlying 
over-approximating abstract 
 domains, which collect constraints on the program 
 variables and, indirectly, on the input data of a 
 program. The abstract domain is 
 parametric in the choice of the underlying domains.

Thus, our analysis 
infers
\emph{necessary 
	conditions} on the data read by the program, 
	i.e., 
conditions such that, if violated, guarantee that the 
program will execute unsuccessfully or 
incorrectly. 
This approach suffers from false 
negatives. However, we argue that this is preferable 
in practice to overwhelming data scientists 
with possibly many false positives (as with 
sufficient conditions).

Back to our motivating example, the analysis 
(parameterized by the sign abstract domain 
\cite{CousotC-92b} and the finite string set domain 
\cite{Christensen-03}) infers 
that 
data files read by the program in 
Figure~\ref{fig:gpa} have the following 
shape:
\begin{equation*}
\begin{array}{rcrclc}
&&&& \scriptstyle{\textcolor{gray}{1}} & 
\textsc{int} \geq 0 \\
\multirow{5}{*}{$d_1$} 
&\multirow{5}{*}{$\begin{cases}
	& \\
	& \\
	& \\
	& \\
	&
	\end{cases}$}
& &&
\scriptstyle{\textcolor{gray}{2}} & 
\textsc{string} \\
&&&& \scriptstyle{\textcolor{gray}{3}} & 
\textsc{int} \geq 0 \\
&&\multirow{2}{*}{$d_3$} 
&\multirow{2}{*}{$\begin{cases}
	& \\
	&
	\end{cases}$}& \scriptstyle{\textcolor{gray}{4}} & 
\textsc{string}\in \set{\text{'A', 'B', 'C', 'D', 
	'F'}} \\
&&&& \scriptstyle{\textcolor{gray}{\vdots}} & 
\dots \\
&&&& \scriptstyle{\textcolor{gray}{\vdots}} & 
 
\end{array}
\end{equation*}
where $d_i$ denotes the data at line $i$ of the 
data file. Thus, the analysis would detect the 
mistake discussed above, since a data file 
containing the erroneous data does not match this 
inferred condition.

Note that, in general, a mismatch between a data 
file and a data-processing program indicates a 
mistake either in data or in the source code of the 
program. Our analysis does not aim to address this 
question. More generally, the result of our analysis 
can be used 
for a wide range of applications: from  
code specifications \cite{Cousot13}, to 
grammar-based 
testing \cite{Hennessy05}, to automatically 
checking and 
guiding the 
cleaning of the 
data \cite{Radwa18,Madelin17}.

%
%
%

\subsubsection{Outline.}
Section~\ref{sec:semantics} introduces the syntax 
and concrete semantics of our data-processing 
programs. In Section~\ref{sec:constraining}, we 
define and present instances of the underlying 
abstract domains. We describe the rest our data 
shape abstract domain in Section~\ref{sec:stack} 
and define the abstract semantics in 
Section~\ref{sec:abstract}. Our prototype static 
analyzer is presented in 
~\ref{sec:implementation}. 
Finally, Section~\ref{sec:related} discusses related 
work and Section~\ref{sec:conclusion} concludes 
and envisions future work.

\section{Input Data-Aware Program Semantics}
\label{sec:semantics}

\subsubsection{Input Data.}

We consider \emph{tabular data} 
stored, e.g., in CSV files. We note, however, that 
what we 
present easily generalizes to other files 
as, 
e.g., spreadsheets. 

Let $\svals$ be a set of string values. 
Furthermore, 
let $\isvals \subseteq \svals$ and 
$\fsvals \subseteq \svals$ be the sets of string 
values that can be interpreted as integer and float 
values, respectively.
We formalize a data file as a possibly empty $(r 
\times c)$-matrix of string values, where $r \in 
\nat$ and $c \in \nat$ denote the number of matrix 
row (i.e., data records)
and columns (i.e., data fields), respectively. We 
write $\epsilon$ to denote an empty data file.
Let 
\begin{equation}\label{eq:files}
\files \defined \bigcup_{r \in \nat}\bigcup_{c \in 
\nat} \svals^{r \times c}
\end{equation}
be the set of all data files. Without loss of 
generality, to simplify our formalization, we assume 
that data 
records contain only one field, i.e., $r = 1$. We lift 
this assumption and consider multiple data fields in 
Section~\ref{subsec:other}.

\subsubsection{Data-Processing 
Language.}

\begin{figure}[t]
	\begin{center}
		\begin{tabular}{lclr}
			A &$\Coloneqq$& $X$ & $X\in 
			\vars$ 
			\\
			&$\vert$& $v$ & $v \in \vals$ \\
			&$\vert$& 
			$\ipt~~\vert~~\mathsf{\mathbf{int}}(A)~~\vert~~\mathsf{\mathbf{float}}(A)~$
			 & \\
			&$\vert$& $A_1 \diamond A_2 $ &
			\hspace{3em}$\diamond \in 
			\set{+,-,*,/}$ 
			\\
			\\
			B & $\Coloneqq$& $A_1 \bowtie A_2$ 
			&
			$\bowtie~\in
			\{<, \leq, =, \not=, >, \geq\}$ \\
			&$\vert$ & $
			B_1 \lor 
			B_2~~\vert~~B_1 
			\land B_2$ \\
			\\
			S & $\Coloneqq$& $^lX := A$ & $l \in 
			\labels, X \in 
			\vars$ \\
			&$\vert$& 
			$\mathsf{\mathbf{if}}~^lB~\mathsf{\mathbf{then}}~S_1~\mathsf{\mathbf{else}}~S_2~\mathsf{\mathbf{fi}}$
			 & $l \in 
			 \labels$ \\
			 &$\vert$& 
			 $\mathsf{\mathbf{for}}~^lA~\mathsf{\mathbf{do}}~S~\mathsf{\mathbf{od}}~~\vert~~\mathsf{\mathbf{while}}~^lB~\mathsf{\mathbf{do}}~S~\mathsf{\mathbf{od}}$
			 & $l \in 
			 \labels$ \\
			 &$\vert$&  $S_1 ; S_2$ & \\
			 \\
			 P & $\Coloneqq$& $S^l$ & $l \in 
			 \labels$
		\end{tabular}
	\end{center}
	\vspace{-1em}
	\caption{Syntax}\label{fig:syntax}
\end{figure}

We consider a toy \python-like
programming language for data 
manipulation, which we use for illustration 
throughout the rest of the paper. Let \vars be a 
finite set of program 
variables, and 
let $\vals\defined\ivals \cup \fvals \cup 
\svals$ be 
a set of values partitioned  in sets of integer 
(\ivals), float (\fvals), and string (\svals) values. 
The 
syntax of programs is defined inductively in 
Figure~\ref{fig:syntax}.
A program $P$ consists of an instruction $S$ 
followed 
by a 
unique label $l \in \labels$. Another unique label 
appears within each instruction.
Programs can read data from an input 
data file: the 
$\ipt$ expression 
consumes a record from the input data file.
 Without loss of generality, to simplify our 
 formalization, we assume that only 
 the right-hand sides of assignments can contain 
 $\ipt$ sub-expressions.
(Programs can always be rewritten to 
satisfy this 
assumption.)
 The 
 $\mathsf{\mathbf{for}}~A~\mathsf{\mathbf{do}}~S~\mathsf{\mathbf{od}}$
 instruction repeats an instruction $S$ for $A$ 
 times.
 The rest of the language syntax is standard. 

\subsubsection{Input-Aware Semantics.}

We can now define the (concrete) semantics of the 
data-processing programs. This semantics differs 
from the usual semantics in that it is 
\emph{input 
data-aware}, that is, it explicitly considers 
the data read by programs.

An environment $\rho\colon \vars \rightarrow 
\vals$ maps each program variable $X \in \vars$ to 
its value $\rho(X) \in \vals$. Let \envs denote the 
set of all environments. 

The semantics of an 
arithmetic expression $A$ is a function 
$\function{\arith{A}}{\envs\times\files}{\vals\times\files}$
 mapping an environment and a data file to the 
 value (in $\vals$) of  the expression in the 
given environment and given the data read from the 
 file (if any), and the (rest of) the data file (in 
 $\files$) after the 
 data is consumed. 
 
\begin{example}\label{ex:arith}
	Let $\rho$ be an environment that 
	maps the variable 
	$gpa$ to the value 
	\text{\lstinline[language=Python]{3.0}}, and let 
	$D 
	= \left[
	\begin{matrix}
	\text{\lstinline[language=Python]{4.0}} \\
	\text{\lstinline[language=Python]{1.0}} \\
	\text{\lstinline[language=Python]{3.0}}
	\end{matrix} \right]$ be a data file containing 
	three 
	data records. 
	We consider the expression $gpa + 
	\ipt$, which simplifies the right-hand side of the 
	assignment at line 9 in Figure~\ref{fig:gpa}.
Its semantics
	is 
	$\arith{
		gpa + \ipt
	} = \left( \text{\lstinline[language=Python]{7.0}}, 
	\left[
	\begin{matrix}
	\text{\lstinline[language=Python]{1.0}} \\
	\text{\lstinline[language=Python]{3.0}}
	\end{matrix} \right] \right)$. \qee
\end{example}
 
We also define the standard input-agnostic 
semantics
$\function{\aritha{A}}{\envs}{\powerset{\vals}}$ 
mapping an environment to the set of all possible 
values of the expression in the environment: 
$\aritha{A}\rho \defined \set{v \in \vals \mid 
\exists D \in \files\colon \tuple{v}{\_} = 
\arith{A}\tuple{\rho}{D}}$.
 
Similarly, the semantics of a 
 boolean 
 expression 
 $\function{\bool{B}}{\envs}{\set{\textsf{tt},
 		\textsf{ff}}}$ maps an environment 
 to the truth value of the expression in the 
 given environment.
%

\begin{figure}[t]
	\begin{align*}
	&\stmt{^lX := A}W \defined \set{
		\tuple{\rho}{D} \in \envs\times\files \mid 
		\tuple{v}{R} = \arith{A}\tuple{\rho}{D}, 
		\tuple{
			\rho\lbrack X \mapsto v \rbrack
		}{
		R
		} \in W
	} \\
	&\stmt{\mathsf{\mathbf{if}}~^lB~\mathsf{\mathbf{then}}~S_1~\mathsf{\mathbf{else}}~S_2~\mathsf{\mathbf{fi}}}W
	 \defined W1 \cup W2 \\
	 &\qquad W1 \defined \set{
	 	\tuple{\rho}{D} \in \envs\times\files \mid 
	 	\textsf{tt} = \bool{B}\rho, 
	 	\tuple{
	 		\rho
	 	}{
	 		D
	 	} \in \stmt{S_1}W
	 } \\
	 &\qquad W2 \defined \set{
	 	\tuple{\rho}{D} \in \envs\times\files \mid 
	 	\textsf{ff} = \bool{B}\rho, 
	 	\tuple{
	 		\rho
	 	}{
	 		D
	 	} \in \stmt{S_2}W
	 }  \\
&\stmt{\mathsf{\mathbf{for}}~^lA~\mathsf{\mathbf{do}}~S~\mathsf{\mathbf{od}}}W
\defined \set{
	\tuple{\rho}{D} \in \envs\times\files \mid 
	\tuple{0}{D} = \arith{A}\tuple{\rho}{D}, 
	\tuple{
		\rho
	}{
		D
	} \in W
} \cup W' \\
	&\qquad W' \defined \set{
		\tuple{\rho}{D} \in \envs\times\files 
		~\left\vert~ 
		\begin{matrix}
		\tuple{v}{D} = \arith{A}\tuple{\rho}{D}, 
		v > 0, \\
		\tuple{
			\rho
		}{
			D
		} \in \stmt{S; 
		\mathsf{\mathbf{for}}~^lA-1~\mathsf{\mathbf{do}}~S~\mathsf{\mathbf{od}}}W
	\end{matrix}
	\right.
	}  \\
&\stmt{\mathsf{\mathbf{while}}~^lB~\mathsf{\mathbf{do}}~S~\mathsf{\mathbf{od}}}W
\defined \lfp~F \\
&\qquad F(Y) \defined \set{
	\tuple{\rho}{D} \in \envs\times\files \mid 
	\textsf{ff} = \bool{B}\rho, 
	\tuple{
		\rho
	}{
		D
	} \in W
} \cup W' \\
&\qquad W' \defined \set{
	\tuple{\rho}{D} \in \envs\times\files \mid 
	\textsf{tt} = \bool{B}\rho, 
	\tuple{
		\rho
	}{
		D
	} \in \stmt{S}Y
} \\
&\stmt{S_1; S_2}W \defined \stmt{S_1} \circ 
\stmt{S_2}W
	\end{align*}
	\caption{Input-Aware Concrete Semantics of 
	Instructions}\label{fig:concrete}
\end{figure}

The semantics of programs 
$\function{\Delta\semantics{P}}{\labels}{\powerset{\envs\times\files}}$
 maps each program 
label $l \in \labels$ to the set of all pairs of
environments 
that are possible when the program execution 
is at that label, and input data 
files 
that the program can 
\emph{fully 
read without errors} 
starting \emph{from} that label. 
We define this semantics \emph{backwards}, 
starting from  
the final program label where all environments in 
\envs are possible
but only the empty data file $\epsilon$ can be read 
from that program label:
\begin{equation}
\Delta\semantics{P} = \Delta\semantics{S^l} 
\defined \Delta\semantics{S}\left(\lambda p. 
\begin{cases} 
\envs\times\set{\epsilon} & p = l \\
\text{undefined} & \text{otherwise}
\end{cases}
\right)
\end{equation}
In Figure~\ref{fig:concrete}, we (equivalently) define 
the semantics 
$\function{\Delta\semantics{S}}{(\labels\rightarrow\powerset{\envs\times\files})}
{(\labels\rightarrow\powerset{\envs\times\files})}$
of each 
instruction pointwise within
$\powerset{\envs\times\files}$: each function 
$\function{\stmt{S}}{\powerset{\envs\times\files}}{\powerset{\envs\times\files}}$
takes as input a set $W$ of pairs of environments 
and data files and outputs the pairs of possible 
environments and data files that can be read from 
the program
label within the instruction $S$.
%

\begin{example}\label{ex:stmt}
	Let $\rho'$ be an environment 
	that 
	maps the variable 
	$gpa$ to the value 
	\lstinline[language=Python]{7.0}, and let 
	$R 
	= \left[
	\begin{matrix}
	\text{\lstinline[language=Python]{1.0}} \\
	\text{\lstinline[language=Python]{3.0}}
	\end{matrix} \right]$ be a data file. 
	We consider the assignment $gpa := gpa + \ipt$ 
	which simplifies the assignment at line 9 in 
	Figure~\ref{fig:gpa}.	
	Its semantics,
	given $W = \set{ \tuple{\rho'}{R} }$,
	is 
	$\stmt{gpa := gpa + \ipt}W = 
	\set{\tuple{\rho}{D}}$ 
	where $\rho$ maps the variable $gpa$ to the 
	value 
	\lstinline[language=Python]{3.0} and  $D 
	= \left[
	\begin{matrix}
	\text{\lstinline[language=Python]{4.0}} \\
	\text{\lstinline[language=Python]{1.0}} \\
	\text{\lstinline[language=Python]{3.0}}
	\end{matrix} \right]$ (see 
	Example~\ref{ex:arith}). \qee
\end{example}

\subsubsection{Data Shape Abstraction.}

In the following sections, we design a decidable  
abstraction of 
$\Delta\semantics{P}$ which 
\emph{over-approximates} 
the concrete semantics of $P$ at each 
program 
label $l \in \labels$. As a consequence, this 
abstraction yields \emph{necessary 
preconditions} for a program to execute 
successfully and correctly. In particular, if a data 
file is not in 
 the abstraction, the program will definitely 
eventually run 
into an error or compute a wrong result
if it tries to read data from it. On the 
other hand, if a data file is in the abstraction there 
is no guarantee that the program will execute 
successfully and correctly when reading data from 
it.


We derive the abstraction 
$\function{\Delta^\natural\semantics{P}}{\labels}{\aelm}$
 by \emph{abstract 
interpretation} 
\cite{CousotC-POPL77}. No approximation is made 
on \labels. On the other hand, each program label 
$l \in \labels$ is 
associated to an element $Q \in \aelm$ of the 
\emph{data shape
abstract 
domain} \adom. $Q$ over-approximates 
the  possible
environments and data files read starting from $l$.

\begin{figure}[t]
\center
\includegraphics[width=0.75\textwidth]{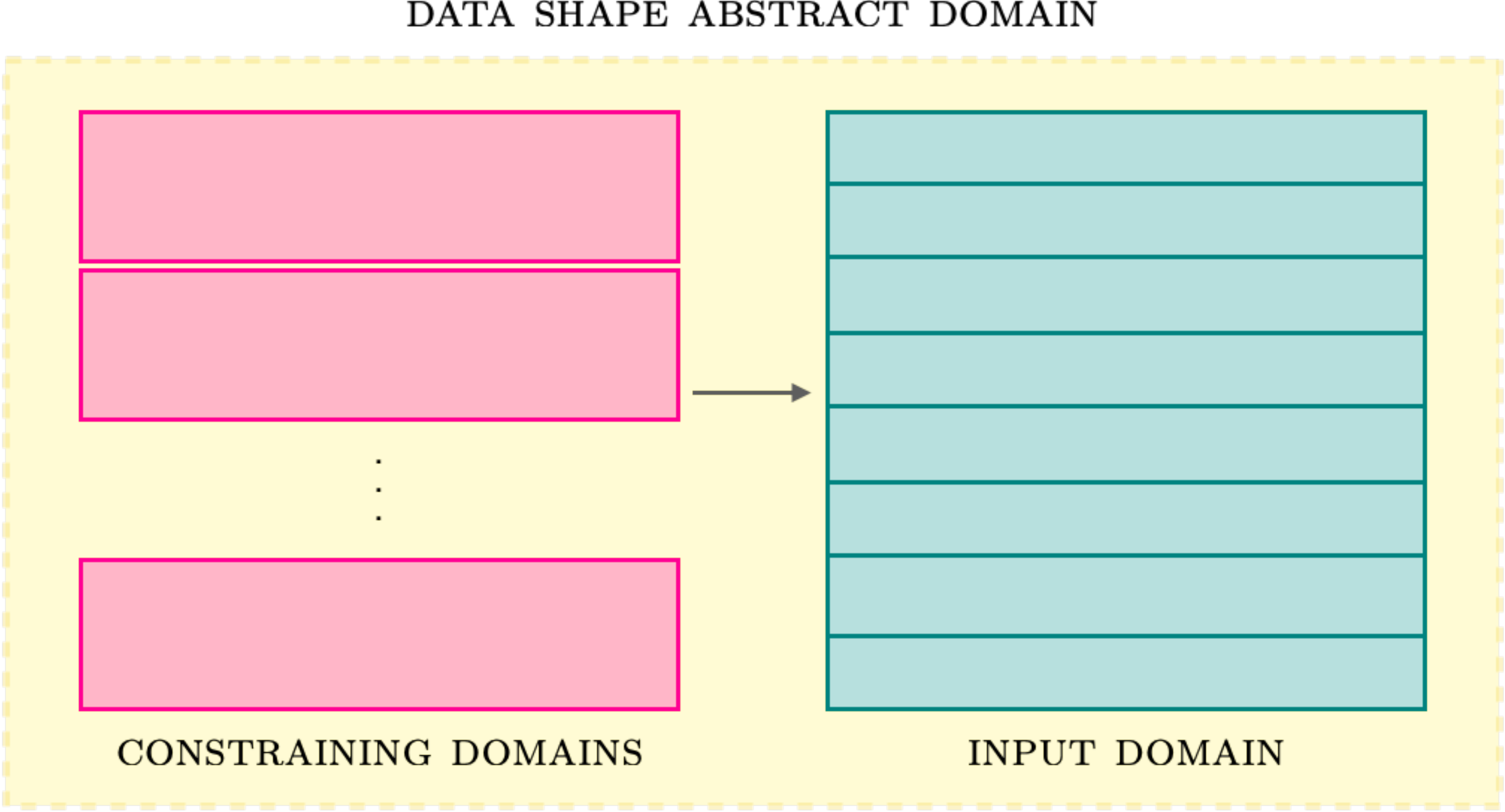}
\caption{Data Shape Abstract 
Domain.}\label{fig:domain}
\end{figure}

An overview of the data shape 
abstract 
domain is given in Figure~\ref{fig:domain}.
It is parameterized by a family $\kdom_1, \dots, 
\kdom_k$ of \emph{constraining abstract 
domains}, which collect constraints on the program 
variables, and an \emph{input abstract domain} 
$\hdom$, which collects constraints on the input 
data read by the program. We now present and 
describe 
instances of these abstract domains, before 
defining $\Delta^\natural\semantics{P}$.

\section{Constraining Abstract 
Domains}\label{sec:constraining}

The constraining abstract domains abstract the 
possible environments at each program label. Thus, 
they constrain the values of the variables of the 
analyzed program and also \emph{indirectly} 
constraint the input data read by the program.

Any constraining domain $\kdom$ that we present 
is characterized by a choice of:
\begin{itemize}
	\item[$\bullet$] a set $\kelm$ of 
	computer-representable abstract domain 
	elements;
	\item[$\bullet$] a partial order 
	$\sqsubseteq_\kdom$ between domain 
	elements;
	\item[$\bullet$] a concretization function 
	$\function{\gamma_\kdom}{\kelm}{\powerset{\envs}}$
	 mapping abstract domain elements to sets of  
	 possible 
	 environments, or, when possible, a 
	Galois connection 
	$\tuple{\powerset{\envs}}{\subseteq} 
	\galois{\alpha_\kdom}{\gamma_\kdom} 
	\tuple{\kelm}{\sqsubseteq_\kdom}$;
	\item[$\bullet$] a least element $\bot_\kdom 
	\in \kelm$ such that 
	$\gamma_\kdom(\bot_\kdom) = \emptyset$;
	\item[$\bullet$] a greatest element $\top_\kdom 
	\in \kelm$ such that 
	$\gamma_\kdom(\top_\kdom) = \envs$;
	\item[$\bullet$] a sound join operator 
	$\sqcup_\kdom$ such that 
	$\gamma_\kdom(K_1) \cup 
	\gamma_\kdom(K_2) \subseteq 
	\gamma_\kdom(K_1 \sqcup_\kdom K_2)$;
	\item[$\bullet$] a sound widening  
	$\triangledown_\kdom$ if $\kdom$ does 
	not satisfy the ascending chain condition;
	\item[$\bullet$] a sound backward assignment
	operator $\assign[\kdom]{X := A}$ 
	such that \\
	$\set{
		\rho \in \envs \mid \exists v \in 
		\aritha{A}\rho\colon \rho[X 
		\mapsto v] \in 
		\gamma(K)} \subseteq 
	\gamma_\kdom(\assign[\kdom]{X := 
	A}K)$; and
\item[$\bullet$] a sound filter operator 
$\filter[\kdom]{B}$ 
such that \\
$\set{ \rho \in \gamma_\kdom(K) \mid \textsf{tt} 
\in \bool{B}\rho
} \subseteq \gamma_\kdom(\filter[\kdom]{B}K)$.
\end{itemize}

Essentially any of the existing classical abstract 
domains 
\cite[etc.]{Costantini-15,CousotC-76,Mine-06} can 
be a constraining 
domain. 
Some of their operators just need to be augmented 
with 
certain
operations to ensure the communication with the 
input domain $\hdom$, which (directly) constraints 
the input 
data.

Specifically, the backward assignment operation 
$\assign[\kdom]{X := A}$ 
needs to be preceded by a $\replace{A, 
\ivars}$ 
operation, which replaces 
each $\ipt$ sub-expressions of 
$A$ with a fresh special 
input variable $I \in 
\ivars$, The input variables are added to the 
constraining domain on the fly to track the value of 
the 
input data as well as the \emph{order} in which 
the 
data is read by the program.

\begin{example}\label{ex:replace}
	Let us consider again the assignment $gpa := gpa + 
	\ipt$ which simplifies line 9 in 
	Figure~\ref{fig:gpa}.
	On way to track the order in which input data is read 
	by the program is to parameterize the fresh input 
	variables by the program label at which the 
	corresponding \ipt expression occur. 
	If we use line numbers as labels, in this case we only 
	need one fresh input variable $I_9$ (for multiple \ipt 
	expressions at the same program label $9$ we can 
	add 
	superscripts: $I^1_9, I^2_9, \dots$). Thus, 
	$\replace{gpa + 
		\ipt, \set{I_9}} = gpa + I_9$. \qee
\end{example}


Once the assignment or filter operation has been 
performed, the operation $\record{I}$ extracts from 
the domain the constraints on each newly added 
input variable $I$ so that they can be directly 
recorded in the input domain $\hdom$. The input 
variables can then be removed from the 
constraining 
domain $\kdom$.

\subsection{Non-Relational Constraining 
Abstract Domains}\label{subsec:nonrel}

In the following, we present a few instances of 
\emph{non-relational} constraining domains. These 
domains
abstract each program variable independently. Thus, 
each constraining domain element $K \in 
\kelm_\uelm$ of 
$\kdom_\udom$ 
is a map 
$\function{K}{\vars}{\uelm}$ from 
program 
variables to elements of a \emph{basis} abstract 
domain $\udom$.

In the following, we write $\uelm\semantics{A}K$ 
to denote the value (in $\uelm$) of an arithmetic 
expression $A$ given the abstract domain element 
$K \in \kelm_\uelm$. In particular, for a binary 
expression $A_1 
\diamond A_2$, we define $\uelm\semantics{A_1 
	\diamond_\udom A_2}K = 
\uelm\semantics{A_1}K \diamond_\udom 
\uelm\semantics{A_2}K$ and thus 
we assume that the basis  
$\udom$ is equipped with the 
operator $\diamond_\udom$.

The concretization 
function 
$\function{\gamma_{\kdom_\udom}}{\kelm_\uelm}{\powerset{\envs}}$
is:
\begin{equation}\label{eq:gamma}
\gamma_{\kdom_\udom}(K) \defined
\set{
	\rho \in \envs \mid
	\forall X \in \vars\colon 
	\mathsf{str}(\rho(X)) \in \gamma_\udom(K(X))
}
\end{equation}
where $\function{\gamma_\udom}{\uelm}{\svals}$ 
and
$\function{\mathsf{str}}{\vals}{\svals}$ 
converts float and integer values to strings 
such that $\mathsf{str}(\fvals) = \fsvals$ and 
$\mathsf{str}(\ivals) = \isvals$.
The partial order $\sqsubseteq_{\kdom}$, 
join 
$\sqcup_{\kdom}$, 
and widening 
$\triangledown_{\kdom}$ are straightforwardly 
defined pointwise. 

For these constraining domains, the 
$\replace{A, \ivars}$ operation \emph{temporarily} 
enlarges the domain of the current abstract element 
$K \in \kelm_\uelm$ to 
also include input 
variables, i.e., 
$\function{K}{\vars\cup\ivars}{\uelm}$. 
The $\record{I}$ operation simply returns the value 
$K(I) \in \uelm$. 
All 
input variable are then 
removed from the domain of $K$.

\subsubsection{Type Constraining Abstract 
Domain.}

The first instance that we consider is very simple 
but interesting to catch exceptions that would be 
raised when casting inputs to integers or floats, as 
at lines 2 and 5 in Figure~\ref{fig:gpa}.

\begin{wrapfigure}{r}{0.4\textwidth}
	\vspace{-25pt}
\begin{center}
	\begin{tikzpicture}
			\node (A) {\textsc{string}};
			\node (B) [below of=A] {\textsc{float}};
			\node (E) [below of=B] {\textsc{int}};
			\node (H) [below of=E] {$\bot_\tdom$};
			\draw (A) -- (B);
			\draw (B) -- (E);
			\draw (E) -- (H);
		\end{tikzpicture}
	\end{center}
	\vspace{-10pt}
	\caption{The $\telm$ type 
	lattice.}\label{fig:type}
	\vspace{-15pt}
\end{wrapfigure}

We define the basis type domain $\tdom$, to 
track \emph{the type of input data that can be 
stored in 
the program 
variables}. Its 
elements belong to the type lattice $\telm$ 
represented by the Hasse 
diagram in Figure~\ref{fig:type}. $\telm$ defines the 
type 
hierarchy (reminiscent of that of \python) that we 
use for our analysis. Data is 
always read as a string (cf. 
Section~\ref{sec:semantics}). Thus, 
\textsc{string} is the highest type in the hierarchy. 
Some (but not all) strings can be cast to float or 
integer, thus the \textsc{float} and \textsc{int} 
types follow in the hierarchy. Finally, $\bot_\tdom$ 
indicates an exception.

We define the concretization function 
$\function{\gamma_\tdom}{\telm}{\svals}$ as 
follows: 
\begin{equation}\label{eq:gammaT}
\begin{array}{cccc}
\gamma_\tdom(\textsc{string}) \defined \svals & 
\hspace{1em}
\gamma_\tdom(\textsc{float}) \defined \fsvals & 
\hspace{1em}
\gamma_\tdom(\textsc{int}) \defined \isvals & 
\hspace{1em}
\gamma_\tdom(\bot_\tdom) \defined
\emptyset
\end{array}
\end{equation}
The partial order $\sqsubseteq_\tdom$, join 
$\sqcup_\tdom$, and meet $\sqcap_\tdom$ 
are 
defined by Figure~\ref{fig:type}. No widening 
$\triangledown_\tdom$ is necessary since the 
basis type domain
$\tdom$ 
is finite.
%

Each element $K \in \kelm_\telm$ of the type 
constraining abstract 
domain $\kdom_\tdom$ is thus a map 
$\function{K}{\vars}{\telm}$ from program 
variables to type elements. 
The bottom element is the constant map $\lambda 
X. \bot_\tdom$ which represent a program 
exception. The top element is $\lambda 
X. \textsc{string}$ or, better, $\lambda 
X. \textsf{type}(X)$, where $\textsf{type}(X)$ is the 
type inferred for $X$ by a static type inference 
previously run on the program (e.g., 
\cite{Hassan-18,Monat-20} for 
\python).
In the latter case, the analysis with $\kdom_\tdom$ 
might refine the inferred type (e.g., $\textsf{type}(X) 
= \textsc{float}$ but the analysis finds 
$K(X) = \textsc{int}$). In particular, such a 
refinement is done by 
the $\assign[\kdom_\tdom]{X := A}$ and 
$\filter[\kdom_\tdom]{B}$ operators.

The $\assign[\kdom_\tdom]{X := A}$ operator
refines the type 
of input data mapped to from the variables that 
appear in the assigned expression $A$. Specifically, 
$\assign[\kdom_\tdom]{X := A}K \defined 
\refine{\replace{A, \ivars}}{K[X \mapsto 
\textsf{type}(X)]}{K(X)}$, 
where the $\function{\textsc{refine}_A}{\kelm 
\rightarrow \telm}{\kelm}$ function is 
defined as follows:
\begin{equation*}
\begin{array}{rclr}
	\refine{X}{K}{T} & \defined & K[X \mapsto 
	K(X) \sqcap_\tdom T] & X \in 
	\vars 
	\\
	\refine{v}{K}{T} & \defined & K & v \in \vals \\
	\refine{I}{K}{T} & \defined & K[I \mapsto T] & I \in 
	\ivars \\
	\refine{\mathsf{\mathbf{int}}(A)}{K}{T} & \defined 
	& \refine{A}{K}{T \sqcap_\tdom \textsc{int}} &  
	\\
	\refine{\mathsf{\mathbf{float}}(A)}{K}{T} & 
	\defined 
	& \refine{A}{K}{T \sqcap_\tdom \textsc{float}} &  
	\\
	\refine{A_1 \diamond A_2}{K}{T} & 
	\defined 
	& \refine{A_1}{\refine{A_2}{K}{T'}}{T'} 
	 &  \hspace{1em} T' = T \sqcap_\tdom 
	 \textsc{float}
\end{array}
\end{equation*}
Note that, for soundness, the current 
value $K(X)$ of the assigned 
variable $X$ must be  
forgotten before the refinement (i.e., $K[X \mapsto 
\textsf{type}(X)]$). We  
refine variables within an 
arithmetic operation $A_1 
\diamond A_2$ to contain data of at most type 
$\textsc{float}$.

\begin{example}[continue from 
Example~\ref{ex:replace}]\label{ex:type}
	Let us consider again the assignment $gpa := 
	gpa + 
	\ipt$ which simplifies line 9 in 
	Figure~\ref{fig:gpa} and let $K$ be an abstract 
	domain element 
	which maps the variable $gpa$ to the type value 
	$\textsc{int}$, while a previously ran type 
	inference has determined that $\textsf{type}(gpa) 
	= \textsc{float}$. We have: 
	\begin{equation*}
	\begin{array}{l}
	\assign[\kdom_\tdom]{gpa := gpa + \ipt}K 
	\defined 
	\refine{gpa + I_9}{K[gpa \mapsto 
		\textsc{float}]}{\textsc{int}} \\
	= \refine{gpa}{\refine{I_9}{K[gpa \mapsto 
			\textsc{float}]}{\textsc{int}}}{\textsc{int}} \\
	= \refine{gpa}{K[gpa \mapsto \textsc{float}][I_9 
	\mapsto \textsc{int}]}{\textsc{int}} 
	= K[I_9 
	\mapsto \textsc{int}][gpa \mapsto \textsc{int}]
	\end{array}
	\end{equation*}
	which indicates that the program 
	expects 
	to read an integer at line $9$. Note that, this is a 
	result of our choice for $K$. Indeed, with $K$ 
	mapping $gpa$ to $\textsc{float}$, we have 
	$\assign[\kdom_\tdom]{gpa := gpa + \ipt}K = 
	K[I_9 
	\mapsto \textsc{float}][gpa \mapsto 
	\textsc{float}]$ (which is what the 
	program in Figure~\ref{fig:gpa} actually expects). 
	\qee
\end{example}

Similarly, the filter operator 
$\filter[\kdom_\tdom]{B}$ is defined as follows:
\begin{equation*}
\begin{array}{rclr}
\filter[\kdom_\tdom]{A_1 
	=
A_2}K 
& \defined & 
\refine{A_1}{\refine{A_2}{K}{\telm\semantics{A_1}K}}{\telm\semantics{A_2}K}
 & 
\\
\filter[\kdom_\tdom]{A_1 
	\not=
	A_2}K 
& \defined & 
K
& 
\\
\filter[\kdom_\tdom]{A_1 
\mathrel{\overline{\mbox{$\bowtie$}\raisebox{1.9mm}{}}}
 A_2}K 
& \defined & 
\refine{A_1}{\refine{A_2}{K}{\textsc{float}}}{\textsc{float}}
 & 
\\
\filter[\kdom_\tdom]{B_1 \lor B_2}K & \defined 
& 
\filter[\kdom_\tdom]{B_1}K \sqcup_{\kdom_\ndom} 
\filter[\kdom_\tdom]{B_2}K & 
\\
\filter[\kdom_\tdom]{B_1 \land B_2}K & \defined 
& 
\filter[\kdom_\tdom]{B_2}K \circ
\filter[\kdom_\tdom]{B_1}K & 
\end{array}
\end{equation*}
where 
$\overline{\mbox{$\bowtie$}\raisebox{1.9mm}{}}
 \in \set{<, \leq, 
>, \geq}$. 

The soundness of the domain operators is 
straightforward: 

\begin{lemma}\label{lm:type}
	The operators of the type constraining  
	domain $\kdom_\tdom$ are sound.
\end{lemma}

\subsubsection{Value Constraining Abstract 
Domains.}

Numerical abstract domains such as 
the interval 
domain \cite{CousotC-76} or the sign domain 
\cite{CousotC-92b} can be used to track \emph{the 
	input data values that can be stored in the 
	program variables}. In particular, the latter is 
useful to catch exceptions raised 
when diving by zero, as at line 10 in 
Figure~\ref{fig:gpa}.

\begin{wrapfigure}{l}{0.4\textwidth}
	\vspace{-30pt}
	\begin{center}
		\begin{tikzpicture}[node distance=1cm]
		\node (A) {$\top_\ndom$};
		\node (B) [below of=A] {$\not= 0$};
		\node (C) [left of=B] {$\leq 0$};
		\node (D) [right of=B] {$\geq 0$};
		\node (E) [below of=B] {$= 0$};
		\node (F) [left of=E] {$< 0$};
		\node (G) [right of=E] {$> 0$};
		\node (H) [below of=E] {$\bot_\ndom$};
		\draw (A) -- (B);
		\draw (A) -- (C);
		\draw (A) -- (D);
		\draw (B) -- (F);
		\draw (B) -- (G);
		\draw (C) -- (E);
		\draw (C) -- (F);
		\draw (D) -- (E);
		\draw (D) -- (G);
		\draw (E) -- (H);
		\draw (F) -- (H);
		\draw (G) -- (H);
		\end{tikzpicture}
	\end{center}
	\vspace{-15pt}
\caption{The $\nelm$ sign 
	lattice.}\label{fig:sign}
\vspace{-45pt}
\end{wrapfigure}

The sign lattice $\nelm$ shown in 
Figure~\ref{fig:sign} represents the elements of the 
basis sign domain $\ndom$. We define the 
concretization function 
$\function{\gamma_\ndom}{\nelm}{\svals}$ as 
follows:
\begin{equation}\label{eq:gammaN}
\begin{array}{rcl}
\gamma_\ndom(\top_\ndom) &\defined& \svals 
\\
\gamma_\ndom(\lhd 0) &\defined& 
\fsvals^{\lhd 0}
\\
\gamma_\ndom(\bot_\ndom) &\defined& 
\emptyset \\
\end{array}
\end{equation}
where $\lhd \in \set{<, \leq, =, \not=, >, \geq}$ 
and $\fsvals^{\lhd 0}$ denotes the set of string 
values that can be interpreted as float values that 
satisfy $\lhd 0$.
The partial order $\sqsubseteq_\ndom$, join 
$\sqcup_\ndom$, and meet $\sqcap_\ndom$ are 
defined by the Hasse diagram in 
Figure~\ref{fig:sign}. Again, no widening 
$\triangledown_\ndom$ is necessary since 
the 
basis domain
$\ndom$ is finite.

Each element $K \in \kelm_\nelm$ of the sign 
constraining abstract 
domain $\kdom_\ndom$ is thus a map 
$\function{K}{\vars}{\nelm}$ from program 
variables to sign elements. 

For this domain, 
the backward 
assignment operator is $\assign[\kdom_\ndom]{X 
:= A}K 
\defined 
\refine{\replace{A, \ivars}}{K[X \mapsto 
	\top_\ndom]}{K(X)}$, 
where $\function{\textsc{refine}_A}{\kelm 
	\rightarrow \nelm}{\kelm}$ is:
\begin{equation*}
\begin{array}{rclr}
\refine{X}{K}{N} & \defined & K[X \mapsto 
K(X) \sqcap_\ndom N] &
\hspace{-35em}X \in \vars  
\\
\refine{v}{K}{N} & \defined & K & 
\hspace{-35em}v \in \vals \\
\refine{I}{K}{N} & \defined & K[I \mapsto N] & 
\hspace{-35em}I \in \ivars \\
\refine{\mathsf{\mathbf{int}}(A)}{K}{N} & = 
& \refine{\mathsf{\mathbf{float}}(A)}{K}{N} \defined 
\refine{A}{K}{N} &  
\\
\refine{A_1 + A_2}{K}{N} & 
\defined 
& \refine{A_1}{
\refine{A_2}{K}{
	N -_\ndom \nelm\semantics{A_1}K
}
}{N -_\ndom 
\nelm\semantics{A_2}K} 
&  \\
\refine{A_1 - A_2}{K}{N} & 
\defined 
& \refine{A_1}{
	\refine{A_2}{K}{
		\nelm\semantics{A_1}K -_\ndom N
}
}{N +_\ndom 
\nelm\semantics{A_2}K} 
&   \\
\refine{A_1 * A_2}{K}{N} & 
\defined 
& \refine{A_1}{
\refine{A_2}{K}{
	N \mathrel{/_\ndom} \nelm\semantics{A_1}K}
}{N \mathrel{/_\ndom} 
\nelm\semantics{A_2}K} 
&  
\\
\refine{A_1 / A_2}{K}{N} & 
\defined 
& \refine{A_1}{K'}{N *_\ndom 
\nelm\semantics{A_2}K} 
&  \\
&& K' = \refine{A_2}{K}{
	\not= 0 \sqcap_\ndom 
	(\nelm\semantics{A_1}K \mathrel{/_\ndom} N)
} &
\end{array}
\end{equation*}
Note that we refine variables in the denumerator 
$A_2$ of a division expression $A_1 \div A_2$ to 
have values different from zero.

\begin{example}\label{ex:sign}
	Let us consider the assignment $result := 
	gpa \mathrel{/} classes$ at line 10 in 
	Figure~\ref{fig:gpa} and let $K$ be an abstract 
	domain element 
	which maps the variables $gpa$ and $result$ to 
	the sign 
	value 
	$\geq 0$ and the variable $classes$ to 
	$\top_\ndom$. 
	We have: 
	\begin{equation*}
	\begin{array}{l}
	\assign[\kdom_\ndom]{result := gpa 
	\mathrel{/} 
	classes}K 
	\defined 
	\refine{gpa / classes}{K[result \mapsto 
		\top_\ndom]}{\geq 0} \\
	= \refine{gpa}{\refine{classes}{K[gpa \mapsto 
			\top_\ndom]}{\not= 0}}{\geq 0} \\
	= \refine{gpa}{K[result \mapsto 
	\top_\ndom][classes 
		\mapsto \not= 0]}{\geq 0} \\
	= K[result \mapsto \top_\ndom][classes 
	\mapsto \not= 0][gpa \mapsto \geq 0]
	\end{array}
	\end{equation*}
	which, in particular, indicates that the program 
	expects the variable $classes$ (read at line 5 in 
	Figure~\ref{fig:gpa}) to have a value 
	different from zero. \qee
\end{example}

Instead, the filter operator 
$\filter[\kdom_\ndom]{B}$ is defined as follows:
\begin{equation*}
\begin{array}{rclr}
\filter[\kdom_\ndom]{A \lhd 0}K & \defined & 
\refine{A}{K}{\lhd 0} & \hspace{-5em}\lhd \in 
\set{<, \leq, =, 
\not=, >, \geq}
\\
\filter[\kdom_\ndom]{A_1 \bowtie A_2}K & \defined 
& 
\filter[\kdom_\ndom]{A_1 - A_2 \bowtie 0}K & A_2 
\not= 0
\\
\filter[\kdom_\ndom]{B_1 \lor B_2}K & \defined 
& 
\filter[\kdom_\ndom]{B_1}K 
\sqcup_{\kdom_\ndom} 
\filter[\kdom_\ndom]{B_2}K & 
\\
\filter[\kdom_\ndom]{B_1 \land B_2}K & \defined 
& 
\filter[\kdom_\ndom]{B_2}K 
\circ
\filter[\kdom_\ndom]{B_1}K & 
\end{array}
\end{equation*}

The soundness of the sign constraining domain 
operators follows 
directly from the soundness of the sign abstract 
domain \cite{CousotC-92b}.

\begin{lemma}\label{lm:sign}
	The operators of the sign constraining  
	domain $\kdom_\ndom$ are sound.
\end{lemma}

\subsubsection{String Constraining Abstract 
Domains.}

Finally, we build a last instance of non-relational 
constraining domain on the finite string set domain 
\cite{Christensen-03}, to track 
\emph{the string data values that can be stored in 
the program variables}.
Other more sophisticated string domains exist 
\cite[etc.]{Arceri19,Costantini-15}. However, even 
this simple domain suffices to catch 
\lstinline[language=Python]{KeyError} exceptions 
that might occur, e.g., at line 9 in 
Figure~\ref{fig:gpa}. 

Each abstract domain element $K \in \kelm_\welm$ 
of the string domain $\kdom_\wdom$ is a map 
$\function{K}{\vars}{\welm}$ from program 
variables to an element $W \in \welm$ of the basis 
domain $\wdom$. Elements of $\wdom$ are 
finite sets of at most $m$ string, or the top 
element $\top_\wdom$ which abstracts larger sets 
of strings, i.e., $\welm \defined 
\powerset{\svals} \cup \set{\top_\wdom}$. In the 
following, we write $\bot_\wdom$ to denote the 
empty string set.
The 
concretization function 
$\function{\gamma_\wdom}{\welm}{\svals}$ is:
\begin{equation}\label{eq:gammaW}
\begin{array}{cc}
\gamma_\wdom(\top_\wdom) \defined \svals & 
\hspace{1em}
\gamma_\wdom(W) \defined W 
\end{array}
\end{equation}
The partial order $\sqsubseteq_\wdom$, join 
$\sqcup_\wdom$,
and meet $\sqcap_\wdom$ 
are 
the set operations $\subseteq$, $\cup$, and 
$\cap$ 
extended to also handle 
$\top_\wdom$: 
\begin{equation*}
\begin{array}{rclr}
W_1 \sqsubseteq W_2 & \Leftrightarrow & 
W_2 = \top_\wdom \lor (W_1 \not= \top_\wdom 
\land W_1 \subseteq W_2) &
\\
W_1 \sqcup_\wdom W_2 & \defined & \begin{cases}
\top_\wdom & W_1 = \top_\wdom \lor W_2 = 
\top_\wdom \lor |W_1 \cup W_2| > m \\
W_1 \cup W_2 & \text{otherwise}
\end{cases} \\
W_1 \sqcap_\wdom W_2 & \defined & 
\begin{cases}
W_1 & W_2 = \top_\wdom \\
W_2 & W_1 = \top_\wdom \\
W_1 \cap W_2 & \text{otherwise}
\end{cases}
\end{array}
\end{equation*}
The widening 
$W_1 \triangledown_\wdom W_2$ yields 
$\top_\wdom$ unless $W_2 \subseteq W_1$ (in 
which case it yields $W_1$).

We can now define the backward assignment 
operator 
$\assign[\kdom_\wdom]{X 
	:= A}K 
\defined 
\refine{\replace{A, \ivars}}{K[X \mapsto 
	\top_\wdom]}{K(X)}$, 
where $\function{\textsc{refine}_A}{\kelm 
	\rightarrow \welm}{\kelm}$ is:
\begin{equation*}
\begin{array}{rclr}
\refine{X}{K}{W} & \defined & K[X \mapsto 
K(X) \sqcap_\wdom W] & X \in 
\vars 
\\
\refine{v}{K}{W} & \defined & K & v 
\in \vals \\
\refine{I}{K}{W} & \defined & K[I \mapsto W] & 
I \in 
\ivars \\
\refine{\mathsf{\mathbf{int}}(A)}{K}{W} & = 
& \refine{\mathsf{\mathbf{float}}(A)}{K}{W} \defined 
\refine{A}{K}{W} &  W = \top_\wdom
\\
\refine{\mathsf{\mathbf{int}}(A)}{K}{W} & = 
& \refine{\mathsf{\mathbf{float}}(A)}{K}{W} \defined 
\refine{A}{K}{\bot_\wdom} & W \not= \top_\wdom
\\
\refine{A_1 \diamond A_2}{K}{W} & 
\defined 
& 
\refine{A_1}{\refine{A_2}{K}{W}}{W}
& W = \top_\wdom \\
\refine{A_1 \diamond A_2}{K}{W} & 
\defined 
& 
\refine{A_1}{\refine{A_2}{K}{\bot_\wdom}}{\bot_\wdom}
& W \not= \top_\wdom 
\end{array}
\end{equation*}
Note that, variables in numerical 
expressions (such as $\mathsf{\mathbf{int}}(A)$, 
$\mathsf{\mathbf{float}}(A)$ or $A_1 \diamond 
A_2$) should not have a specific string value (i.e, a 
value different from $\top_\wdom$).

\begin{example}\label{ex:string}
	Let us consider a small extension of our toy 
	language with dictionaries. In particular, we 
	extend the grammar of arithmetic expressions  
	with 
	dictionary display (in \python 
	terminology) expressions $\set{v_0: v_1, v_2: 
	v_3, \dots}$, $v_0, v_1, v_2, v_3, \ldots \in 
	\vals$, for dictionary 
	creation (cf. line 1 in 
	Figure~\ref{fig:gpa}) 
	and dictionary access 
	expressions $X[A]$ (such as 
	$grade2gpa[grade]$ 
	at line 9 in 
	Figure~\ref{fig:gpa}).

	For each dictionary, 
	we assume that abstract domains only keep track 
	of two summary variables \cite{Gopan04}, one 
	representing the 
	dictionary keys and one representing its values. 
	For instance, let us consider the $grade2gpa$ 
	dictionary in Figure~\ref{fig:gpa} and let the 
	string domain element $K$ map 
	the variable $keys(grade2gpa)$ to the set of 
	strings $\set{\text{'A', 'B', 'C', 'D', 'F'}}$ and 
	$values(grade2gpa)$ to $\top_\ndom$. 
	
	We can extend $\textsc{refine}_A$ defined above 
	to 
	handle dictionary access expressions as follows:
	$
	\refine{X[A]}{K}{W} \defined 
	\refine{A}{K}{K(keys(X))}
	$. No refinement can be made on $X$ since, for 
	soundness, only weak updates are allowed on 
	summary variables \cite{Chase90}. 
		For the assignment $gpa := gpa + 
grade2gpa[grade]$ 
at line 9 in Figure~\ref{fig:gpa} 
we thus have $\assign[\kdom_\wdom]{gpa := gpa 
+ grade2gpa[grade]}K = K[grade \mapsto \set{'A', 
'B', 'C', 'D', 'F'}]$, which indicates the string values 
expected by the program for the variable $grade$ 
(read at line 8 in Figure~\ref{fig:gpa}). \qee
\end{example}

The filter operator 
$\filter[\kdom_\wdom]{B}$ is defined as follows:
\begin{equation*}
\begin{array}{rclr}
\filter[\kdom_\wdom]{A_1 
=
	A_2}K 
& \defined & 
\refine{A_1}{\refine{A_2}{K}{\welm\semantics{A_1}K}}{\welm\semantics{A_2}K}
 & 
\\
\filter[\kdom_\wdom]{A_1 
	\not=
	A_2}K 
& \defined & 
K
& 
\\
\filter[\kdom_\wdom]{A_1 
\mathrel{\overline{\mbox{$\bowtie$}\raisebox{1.9mm}{}}}
	A_2}K 
& \defined & 
\refine{A_1}{\refine{A_2}{K}{\welm\semantics{A_1}K}}{\welm\semantics{A_2}K}
 & 
\\
&&&
\hspace{-50em}\welm\semantics{A_2}K = 
\top_\wdom \land \welm\semantics{A_1}K = 
\top_\wdom
\\
\filter[\kdom_\wdom]{A_1 
	\mathrel{\overline{\mbox{$\bowtie$}\raisebox{1.9mm}{}}}
	A_2}K 
& \defined & 
\refine{A_1}{\refine{A_2}{K}{\bot_\wdom}}{\bot_\wdom}
& 
\\
&&&
\hspace{-15em}\welm\semantics{A_2}K \not= 
\top_\wdom \lor \welm\semantics{A_1}K \not= 
\top_\wdom
\\
\filter[\kdom_\wdom]{B_1 \lor B_2}K & 
\defined 
& 
\filter[\kdom_\wdom]{B_1}K 
\sqcup_{\kdom_\wdom} 
\filter[\kdom_\wdom]{B_2}K & 
\\
\filter[\kdom_\wdom]{B_1 \land B_2}K & \defined 
& 
\filter[\kdom_\wdom]{B_2}K 
\circ
\filter[\kdom_\wdom]{B_1}K & 
\end{array}
\end{equation*}
where 
$\overline{\mbox{$\bowtie$}\raisebox{1.9mm}{}}
\in \set{<, \leq, 
	>, \geq}$. 

The soundness of the string constraining domain 
operators follows 
directly from the soundness of the finite string set 
abstract 
domain \cite{Christensen-03}.

\begin{lemma}\label{lm:sign}
	The operators of the string constraining  
	domain $\kdom_\wdom$ are sound.
\end{lemma}

\subsection{Other Constraining Abstract 
Domains}\label{subsec:other}

We now briefly discuss other instances of 
constraining domain. 

\subsubsection{Relational Constraining Abstract 
Domains.}

Other constraining domain can be 
built on \emph{relational} abstract domains. 
Popular such domains are octagons \cite{Mine-06} 
or polyhedra 
\cite{CousotH-POPL78}, which track linear 
relations between program variables.

We refer to the literature for the formal 
definition of these abstract domains and only 
discuss 
here the implementation of the additional 
operations needed to 
communicate with the input domain \hdom.
In particular, similarly to non-relational domains, 
the $\replace{E, \ivars}$ operation temporarily adds 
the input variables in $\ivars$ to the current 
abstract element $K \in \kelm$. These are 
unconstrained at first and might become subjects 
to constraints after an assignment or filter 
operation.

The implementation of the $\record{I}$ operation is 
more complex for relational domains: $\record{I}$
 extracts from the current 
abstract element $K$, an abstract domain 
element $\overline{K}$ containing
all and only the 
constraints in $K$ that involve the input variable 
$I$. The domain $\dom(\overline{K})$ of 
$\overline{K}$ is the subset of $\dom(K)$ 
containing only the variables appearing in these 
constraints. The input variables can then 
be projected away from $K$. 

\begin{example}
	Let us consider again the assignment $gpa := 
	gpa + \ipt$ which simplifies line 9 in 
	Figure~\ref{fig:gpa} and 
	let $K = \set{gpa \geq 0, grades > 0}$ be a 
	polyhedra defined over the variables $gpa$ and 
	$grades$, i.e., $\dom(K) = 
	\set{gpa, grades}$.
	After $\replace{gpa + \ipt, \set{I_9}}$ (cf. 
	Example~\ref{ex:replace}), $K$ is unchanged but 
	its domain is enlarged to also include the input 
	variable $I_9$, i.e., $\dom(K) = 
	\set{gpa, grades, I_9}$. The result of the 
	(replaced) assignment $gpa := gpa + I_9$ is then 
	the polyhedra 
	$K' = \set{gpa + I_9 \geq 0, grades > 0}$.
	Finally, the $\record{I_9}$ operation returns the 
	polyhedra 
	$\overline{K} = 
	\set{gpa + I_9 \geq 0}$, where 
	$\dom(\overline{K}) = \set{gpa, I_9}$.
	\qee
\end{example}

In the following, we assume that input variables are 
parameterized by the program label at which their 
corresponding \ipt expressions occur, as in 
Example~\ref{ex:replace}.
Note that, there is not necessarily a one-to-one 
correspondence between $\ipt$ expressions in a 
program and data record in a data file. Indeed, 
multiple records can be read by the same $\ipt$ 
expression (i.e., in a for loop as in 
Figure~\ref{fig:gpa}) or, vice versa, the same data 
record could be read by multiple $\ipt$ expressions 
(i.e., in different if branches).
%
In particular, 
the latter case implies that two 
abstract domain elements $K_1$ and 
$K_2$ might 
be defined over different input variables. Thus, 
relational constraining domains need to be 
equipped 
with a 
unification operation $\unify{K_1}{K_2}$ to match 
different input variables that correspond to the 
same data record. 
One simple option to deal with this problem is to 
keep track of the order in which 
the input variables are added to a domain element 
by each $\replace{E, \ivars}$ operation. 
The $\unify{K_1}{K_2}$ operation then simply 
consists in matching input variables in their order.

\subsubsection{Container Constraining Abstract 
Domains.}

We now lift the assumption that data records only 
have one field (cf. Section~\ref{sec:semantics}). We 
extend the grammar of expressions to also include 
data access expressions $X[A]$, 
$X \in \vars$, 
(similarly to what we did in Example~\ref{ex:string} 
for dictionaries). Similarly, we extend the grammar 
of statements to also include assignments of the 
form $X[A_1] := A_2$. We call variables like the 
$X$ 
we used in these expressions,
\emph{array variables}.

In this case, abstract domains should be able to also
handle reads and updates of array variables  
in 
addition 
to numerical and 
string variables as so far. The most basic option to 
do so is to use summarization \cite{Gopan04} (as 
in Example~\ref{ex:string}) and only perform weak 
updates \cite{Chase90}. It is sometimes possible to 
fully expand array variables to improve precision 
\cite{Blanchet02}, or use a combination of 
expansion and summarization (i.e., expand part of 
the array up to a certain size and summarize the 
rest).

Many other abstract domains exist that are 
specifically designed to analyze  
arrays 
\cite[etc.]{Cousot11,Gopan05,Halbwachs08,Liu17} 
or, more 
generally, containers (e.g., 
sets, dictionaries) 
\cite[etc.]{Cox13,Cox14,Dillig10,Dillig11,Fulara12}.
Any of these can be instantiated as a constraining 
domain (as we showed in this section) and used 
within 
our framework.

\section{Input Abstract Domain}\label{sec:stack}

The input abstract domain $\hdom$, as mentioned, 
\emph{directly}
constrains the input data read by a program. An 
element $H \in \helm$ of $\hdom$ is 
a 
\emph{stack} of mutable length 
$h$:
\begin{equation*}
\begin{array}{cr}
	R_0 \mid R_1 \mid \dots \mid R_{h-1} \mid 
	R_{h} & \hspace{7em} R_i \in \relm
\end{array}
\end{equation*}
of assumptions on (part of) the input data, or the 
special element $\bot_\adom$ or 
$\top_\adom$. The top element $\top_\adom$ 
denotes unconstrained input data, while 
$\bot_\adom$ 
indicates a program exception.
A 
stack element 
grows 
or shrinks based on the level 
of nesting of the currently analyzed \ipt expression.

Each layer $R_i \in \relm$ is a list of 
$r$ 
assumptions repeated $M$ times:
$\relm \defined \set{ M \cdot (J_i)^r_{i=1} \mid  
J_i 
\in \celm \cup \set{\bigstar} \cup 
\relm}.
$
The \emph{multiplier} $M$ follows this 
grammar:
\begin{equation*}
\begin{array}{lr}
M \Coloneqq X \in \vars 
\mid 
I \in \ivars 
\mid v \in \ivals 
\mid M_1 \diamond M_2 & 
\hspace{5em} \diamond \in 
\set{+,-,*,/}
\end{array}
\end{equation*}
while an assumption $J_i$ can be a \emph{basic 
assumption} in $\celm$, the \emph{dummy 
assumption} 
$\bigstar$, or another list of repeated assumptions 
in $\relm$.

A basic assumption $C \in \celm$ is a family of 
constraints, one for each constraining domain 
$\kdom_1, \dots, \kdom_k$ in $\adom$, 
associated to a particular program label $l \in 
\labels$:
$\celm \defined \set{ \tuple{l}{(Y_i)^k_{i=1}} \mid l 
\in \labels,  Y_i 
	\in \overline{\kelm_i}}$,
where $\overline{\kelm_i} = \uelm_i$ if 
$\kdom_i$ is a non-relational domain, or 
$\kelm_i$ 
otherwise (cf. Section~\ref{sec:constraining}).

\begin{example}\label{ex:basic}
	Let us consider the assignment $grade := I_8$ 
	where $I_8$ is the result of $\replace{\ipt, 
	\set{I_8}}$ at line 8 in 
	Figure~\ref{fig:gpa}. Moreover, let 
	$K_T \in \kdom_\tdom$ and $K_W \in 
	\kdom_\wdom$ map the variable $grade$ to 
	$\textsc{string}$ and $\set{\text{'A', 'B', 'C', 
	'D', 'F'}}$, respectively. After the analysis of the 
	assignment, we have $K_T(I_8) = 
	\textsc{string}$ and $K_W(I_8) = 
	\set{\text{'A', 'B', 'C', 
	'D', 'F'}}$. The call to the function 
	$\record{I_8}$ in the two constraining domains 
	$\kdom_\tdom$ and $\kdom_\wdom$
	effectively creates the basic assumption 
	$\tuple{l_8}{\left[\textsc{string}, \set{\text{'A', 
	'B', 
	'C', 
			'D', 'F'}}\right]}$ in the input domain 
			$\hdom$. 
			\qee	
\end{example}

A repeated assumption $M \cdot (J_i)^r_{i=1}$  
 constrains \emph{all} data read by a for loop.

\begin{example}[continue from 
Example~\ref{ex:basic}]\label{ex:repeated}
	Let us consider the for loop at lines 7-9 in 
	Figure~\ref{fig:gpa}. 
	The \ipt expression at line 8
	is constrained 
	by the basic assumption 
	$\tuple{l_8}{\left[\textsc{string}, \set{\text{'A', 
	'B', 
			'C', 
			'D', 'F'}}\right]}$. Thus, all data read by the 
			for loop is constrained by 
	$classes \cdot \left[
	\tuple{l_8}{\left[\textsc{string}, \set{\text{'A', 'B', 
			'C', 
			'D', 'F'}}\right]}
	\right]$. \qee
\end{example}

Finally, data read by a while loop is generally 
approximated by the dummy assumption 
$\bigstar$, which denotes an unknown number of 
unconstrained data records.

The concretization function 
$\function{\gamma_\hdom}{\helm}{\powerset{\files}}$
 is defined as follows:
 \begin{equation}
 \begin{array}{rcl}
 \gamma_\hdom(\bot_\hdom) & \defined & 
 \emptyset \\
 \gamma_\hdom(H) & \defined & \set{
D \in \files 
\mid 
D \models H
} \\
 \gamma_\hdom(\top_\hdom) & \defined & \files 
 \end{array} 
 \end{equation}
 In particular, the concretization of a stack element 
 $H \in \hdom$ is the set of data files that 
 \emph{satisfy} the assumptions fixed by the 
 stack element. We omit the formal definition of the 
 satisfaction relation $\models$ due to space 
 limitations. The following example should provide 
 an intuition:
 
 \begin{example}\label{ex:gamma}
	Let us assume that the program in 
	Figure~\ref{fig:gpa} is analyzed with $\adom$ 
	instantiated with the type 
	$\kdom_\tdom$, sign $\kdom_\ndom$, and 
	string $\kdom_\wdom$ constraining domains. 
	Let us consider the following stack element $H 
	\in \hdom$ at line 
	$5$:
	\begin{equation*}
		1 \cdot \left[
		\langle l_5, \left[\textsc{int}, \not= 0, 
		\top_\wdom
		\right] \rangle, 
		I_5 \cdot 
		\left[
		\tuple{l_8}{\left[\textsc{string}, \top_\ndom, 
		\set{\text{'A', 'B', 
				'C', 
				'D', 'F'}}\right]}
		\right]
		\right] \mid 1 \cdot \left[\right]
	\end{equation*}	
The data file $\left[
\begin{matrix}
\text{\lstinline[language=Python,mathescape]!$2$!} 
\\
\text{\lstinline[language=Python]{A}} \\
\text{\lstinline[language=Python]{F}}
\end{matrix} \right]$
satisfies $H$ since 
$
\text{\lstinline[language=Python,mathescape]!$2$!}
\in \gamma_\tdom(\textsc{int}) \cap 
\gamma_\ndom(\not= 0) \cap
\gamma_\wdom(\top_\wdom)$ and, similarly, 
$\text{\lstinline[language=Python]{A}}, 
\text{\lstinline[language=Python]{F}} \in 
\gamma_\tdom(\textsc{string}) \cap 
\gamma_\ndom(\top_\ndom) \cap
\gamma_\wdom(\set{\text{'A', 'B', 'C', 'D', 'F'}})$. 
Moreover, 
$I_5 = 
\text{\lstinline[language=Python,mathescape]!$2$!}$
and, indeed, there are exactly two data records 
following 
$\text{\lstinline[language=Python,mathescape]!$2$!}$.
 Instead, the data file $\left[
\begin{matrix}
\text{\lstinline[language=Python,mathescape]!$1$!} 
\\
\text{\lstinline[language=Python]{A}} \\
\text{\lstinline[language=Python]{F}}
\end{matrix} \right]$ (cf. the motivating example in 
the Introduction) does not satisfy $H$ since $I_5 = 
1$ is followed by two data records instead of one. 
\qee
 \end{example}

Any data file satisfies the dummy assumption 
$\bigstar$. Thus any stack element starting with 
the dummy assumption (e.g, $1 \cdot 
\left[\bigstar\right])$ is equivalent to 
$\top_\hdom$.

We define the partial order $\sqsubseteq_\hdom$ 
such that $H_1 \sqsubseteq_\hdom H_2$  
only if 
$\gamma_\hdom(H_1) \subseteq 
\gamma_\hdom(H_2)$. Thus, $H_1 
\sqsubseteq_\hdom H_2$ is always true if 
$H_1 = \bot_\hdom$ or $H_2 = \top_\hdom$. 
Otherwise, $H_1$ and $H_2$ must have the same 
number of layers to be comparable and 
$\sqsubseteq_\hdom$ is defined
laywer-wise. 
Specifically, for each $R_1 = M_1 
\cdot \left[J^1_1, \dots, J^1_{r_1} \right] \in \relm$ 
and $R_2 = M_2 
\cdot \left[J^2_1, \dots, J^2_{r_2} \right] \in 
\relm$, $R_1 \sqsubseteq_\rdom R_2$ if and only 
if $M_1 = M_2$ and $r_1 = r_2$ (i.e., $R_1$ and 
$R_2$ consists of the same number of assumptions 
repeated the same number of times), and 
$\bigwedge^{r_1=r_2}_{i=1} J^1_i 
\sqsubseteq_\mathbb{J} J^2_i$, i.e., $R_1$ imposes 
stronger constraints on the input data than $R_2$. 
The partial order $J_1 \sqsubseteq_\mathbb{J} J_2$  
is again $J_1 \sqsubseteq_\rdom$, if $J_1, J_2 \in 
\relm$. Otherwise, $J_1 \sqsubseteq_\mathbb{J} 
J_2$ is
always true when $J_2 = \bigstar$. For 
basic assumptions $J_1 
= \tuple{l_1}{\left[
	Y^1_0, \dots, Y^1_k
	\right]} \in \celm$ and 
$J_2 = \tuple{l_1}{\left[
	Y^2_0, \dots, Y^2_k
	\right]} \in \celm$, $J_1 \sqsubseteq_\mathbb{J} 
	J_2$ is true if and only if $\bigwedge^k_{i_1} Y^1 
	\sqsubseteq_{\overline{\kdom}_i} 
	Y^2$, where $\overline{\kdom}_i = \udom_i$ if 
	$\kdom_i$ is a non-relational domain, or 
	$\kdom_i$ otherwise. Note that, for relational 
	domains, a unification must be performed prior 
	to $\sqsubseteq_\mathbb{J}$ as discussed in  
	 Section~\ref{sec:constraining}. No comparison is 
	 possible when $J_1 \in \celm$ and $J_2 \in 
	 \relm$, or vice versa.

This is a rather rigid definition for 
$\sqsubseteq_\hdom$. Indeed, in some cases, 
$H_1 \not\sqsubseteq_\hdom H_2$ even though 
$\gamma_\hdom(H_1) \subseteq 
\gamma_\hdom(H_2)$, e.g., consider $H_1 = 1 
\cdot \left[
\tuple{l_a}{\left[\textsc{int}\right]}, 
\tuple{l_b}{\left[\textsc{float}\right]}
\right]$ and $H_2 = 2 \cdot \left[
\tuple{l_c}{\left[\textsc{float}\right]}
\right]$.
Such incomparable stack elements may result from 
syntactically different but semantically close 
programs \cite{DelmasM19} (e.g., in one program a 
loop has been unrolled but not in the other), but 
never during the analysis of a single program. Thus, 
for our purposes, this definition of 
$\sqsubseteq_\hdom$ suffices.

The join $\sqcup_\hdom$
is defined analogously to 
$\sqsubseteq_\hdom$. We omit 
its formal definition due to space limitations.
The join of incomparable stack layers is 
approximated with the dummy layer 
$1 \cdot \left[\bigstar\right]$. Thus, no widening 
$\triangledown_\hdom$ 
is 
needed. 

%

The backward assignment operator 
$\assign[\hdom]{X := A}$ and filter operator 
$\filter[\hdom]{B}$ operate on each 
stack layer independently. For each $R = M \cdot 
(J_i)^r_{i=1} \in \relm$, the assignment replaces 
any occurrence of $X$ in the multiplier $M$ with 
the expression $\replace{A, 
	\ivars}$.
The assignment (resp. filter) operation is done
recursively
on each assumption $J_i$. When $J_i \in \celm$, 
the assignment (resp. filter) is delegated to the  
constraining domains directly. 

\begin{example}[continue from 
Example~\ref{ex:repeated}]\label{ex:assign}
	Let us consider again the assumption $classes 
	\cdot \left[
	\tuple{l_8}{\left[\textsc{string}, \set{\text{'A', 'B', 
			'C', 
			'D', 'F'}}\right]}
	\right]$, which constrains the data read by the 
	for loop at lines 7-9 in 
	Figure~\ref{fig:gpa}, and 
	the assignment $classes := \ipt$ (cf. line 5). The 
	assignment simply replaces the multiplier 
	$classes$ in 
	the assumption with the input variable $I_5$: 
	$I_5 
	\cdot \left[
	\tuple{l_8}{\left[\textsc{string}, \set{\text{'A', 'B', 
			'C', 
			'D', 'F'}}\right]}
	\right]$. \qee
\end{example}

During the analysis of a for loop, the $\rpt{A}$ 
operator modifies the multiplier of the assumption 
in the first stack layer: $\rpt{A}(M \cdot [J, \dots] 
\mid \dots \mid R_h) \defined (A * M) \cdot [J, 
\dots] \mid \dots \mid R_h$. The resulting 
multiplier expression is then simplified, whenever 
possible (e.g, $(X + 1) - 1$ is simplified to $X$).

Finally, it remains to discuss how stack elements 
$H \in \helm$ grow and shrink during the analysis 
of a program. Whenever the analysis enters the 
body of an if or loop statement, the $\push(H)$ 
operation simply adds an extra layer to $H$ 
containing the 
empty assumption $1 \cdot \left[\right]$: 
$\push(H) \defined 1 \cdot \left[\right] \mid H$.
When the analysis later leaves the body of the 
statement, the $\pop(H)$ operation inserts the 
assumption in the first layer into the assumption 
in the second layer: $\pop(R_0 \mid M \cdot [J, 
\dots] \mid \dots \mid R_h) = M \cdot [R_0, J, 
\dots] 
\mid \dots \mid R_h$. Instead, the 
$\overline{\pop}$ operation merges the assumption 
in the first layer with the (first) assumption in the 
second layer: $\pop(R_0 \mid M \cdot [J, 
\dots] \mid \dots \mid R_h) = M \cdot [R_0 
\sqcup_\mathbb{J} J, 
\dots] 
\mid \dots \mid R_h$.

The input domain operators ultimately build on the 
operators of the constraining domains. Thus, their 
soundness directly follows from that of the 
constraining 
domain operators.

\begin{lemma}\label{lm:sign}
	The operators of the input 
	domain $\hdom$ are sound.
\end{lemma}

\section{Input Data-Aware Program 
Abstraction}\label{sec:abstract}

\begin{figure}[t]
	\begin{align*}
	&\astmt{^lX := A}Q \defined \assign[\adom]{X := 
	A}Q \\
	&\astmt{\mathsf{\mathbf{if}}~^lB~\mathsf{\mathbf{then}}~S_1~\mathsf{\mathbf{else}}~S_2~\mathsf{\mathbf{fi}}}Q
	\defined Q_1 \sqcup_\adom Q_2 \\
	&\qquad Q_1 \defined \pop \circ 
	\filter[\adom]{B} \circ 
	\astmt{S_1} \circ \push(Q) \\	
	&\qquad Q_2 \defined \pop \circ 
	\filter[\adom]{\neg B} 
	\circ 
	\astmt{S_2} \circ \push(Q) \\	
	&\astmt{\mathsf{\mathbf{for}}~^lA~\mathsf{\mathbf{do}}~S~\mathsf{\mathbf{od}}}Q
	\defined \lfp^\natural_{\pop \circ \rpt{A} \circ 
	\astmt{S} \circ \push(W)}~G \\
	&\qquad G(Y) \defined 
	\overline{\pop} \circ \rpt{A} \circ \astmt{S} \circ 
	\push(Y)	
	\\	
	&\astmt{\mathsf{\mathbf{while}}~^lB~\mathsf{\mathbf{do}}~S~\mathsf{\mathbf{od}}}Q
	\defined \lfp^\natural~F \\
	&\qquad F(Y) \defined \pop \circ 
	\filter[\adom]{\neg 
	B} \circ \push(Q) \sqcup_\adom \pop \circ 
	\filter[\adom]{B} \circ \astmt{S} \circ \push(Y) \\
%
	&\astmt{S_1; S_2}Q \defined 
	\astmt{S_1} \circ \astmt{S_2}Q
	\end{align*}
	\caption{Input-Aware Abstract Semantics of 
		Instructions}\label{fig:abstract}
\end{figure}

We can now use the data shape abstract domain 
$\adom$ to define the abstract semantics 
$\Delta^\natural\semantics{P}$. We write 
$\tuple{\langle K_1, \dots, K_k \rangle}{H} \in 
\aelm$ to denote an element of $\adom$, where 
$K_1 \in \kelm_1, \dots, K_k \in \kelm_k$ are 
elements of the constraining domains $\kdom_1, 
\dots, \kdom_k$ and $H 
\in \helm$ is an element of the input domain. The 
abstract data shape semantics of a data-processing 
program $P$ is thus:
\begin{equation}
\Delta^\natural\semantics{P} = 
\Delta^\natural\semantics{S^l} 
\defined \Delta^\natural\semantics{S}\left(\lambda 
p. 
\begin{cases} 
\tuple{\langle \top_{\kdom_1}, \dots, 
	\top_{\kdom_k} \rangle}{1 \cdot 
	\left[\right]} & p = l \\
\text{undefined} & \text{otherwise}
\end{cases}
\right)
\end{equation}
The semantics
$\function{\Delta^\natural\semantics{S}}{(\labels\rightarrow\aelm)}
{(\labels\rightarrow\aelm)}$ of each instruction is 
(equivalently) defined pointwise within $\aelm$ in 
Figure~\ref{fig:abstract}: each function 
$\function{\astmt{S}}{\aelm}{\aelm}$ 
over-approximates the possible environments and 
data files that can be read from the program label 
within the instruction $S$. The $\assign[\adom]{X 
:= A}$ operator first invokes 
$\assign[\kdom_i]{X 
:= A}$ on each constraining domain $\kdom_i$. 
Then, the $\record{I}$ operation is  
executed for each input variable $I \in \ivars$ 
corresponding to an $\ipt$ sub-expression of $A$. 
Finally, the assignment is performed on the input 
domain by $\assign[\hdom]{X 
	:= A}$. Similarly, the $\filter[\adom]{B}$ 
	operation is first executed on each constraining 
	domain $\kdom_i$ by $\filter[\kdom_i]{B}$, and 
	then on the input domain by $\filter[\hdom]{B}$. 
	The $\rpt{A}$, $\push$, $\pop$, 
	$\overline{\pop}$ have no effect on the 
	constraining domains and only modify the input 
	domain (cf. Section~\ref{sec:stack}).

The abstract semantics of each instruction is sound:
\begin{lemma}
	$\stmt{\gamma_\adom(Q)} \subseteq 
	\gamma_\adom(\astmt{Q})$
\end{lemma}
where the concretization function 
$\function{\gamma_\adom}{\aelm}{\powerset{\envs\times\files}}$
  is $\gamma_\adom(\tuple{\langle K_1, 
  \dots, K_k 
\rangle}{H}) \defined \set{
\tuple{\rho}{D} \in \envs\times\files \mid
\rho \in \gamma_{\kdom_1}(K_1) \cap 
\dots \cap \gamma_{\kdom_1}(K_k), 
D \in \gamma_\hdom(H) 
}$.

Thus, the abstract data shape semantics 
$\Delta^\natural\semantics{P}$ is also sound:

\begin{theorem}
	For each data-processing program $P$, we have 
	$\Delta\semantics{P} \subseteq 
	\gamma_\adom(\Delta^\natural\semantics{P})$.
\end{theorem}

\section{Implementation}\label{sec:implementation}

We have implemented our input data shape 
analysis in the open-source prototype static 
analyzer 
\tool\footnote{\url{https://github.com/caterinaurban/Lyra}}.
 The implementation is in 
\python and, at the time of writing, accepts data 
processing
programs written in a subset of \python without 
user-defined classes. 
Programs are expected to be type-annotated, either 
manually or by a type inference \cite{Hassan-18}.

For the analysis, various constraining domains are 
available: in addition to the \emph{type}, 
\emph{sign}, and \emph{string} 
domains 
presented in Section~\ref{subsec:nonrel}, \tool is 
equipped with the \emph{character inclusion} 
domain \cite{Costantini-15}, as well as the 
\emph{intervals} \cite{CousotC-76}, 
\emph{octagons} \cite{Mine-06}, and 
\emph{polyhedra} domains \cite{CousotH-POPL78}, 
which build upon the \textsc{apron} library 
\cite{JeannetM-CAV09}. 
A native 
(non-\textsc{apron}-based) implementation of the 
intervals domain is also available.
For containers 
(e.g., lists, sets, dictionaries, \dots), a 
summarization-based abstraction \cite{Gopan04} is 
the default. Lists, tuples, and dictionaries can be 
expanded up to a fixed bound beyond which they 
are summarized (cf. 
Section~\ref{subsec:other}). 

The data shape analysis is performed backwards on 
the control 
flow graph of the program with a standard worklist 
algorithm \cite{NielsonNH-99}, using widening at 
loop heads to enforce termination. The precision of 
the analysis can be improved by running a forward 
pre-analysis which collects values information 
about the program variables (e.g., in 
Figure~\ref{fig:gpa}, this would allow the data 
shape analysis to know the values of the keys of 
the $grade2gpa$ dictionary already at line $9$ even 
if the dictionary is not created until line $1$, cf. 
Example~\ref{ex:string}).

\tool outputs the analysis results in \textsc{json} 
format so that other applications (e.g., automated 
data checking tools \cite{Radwa18,Madelin17}) can 
easily 
interface with it. 

Below, we 
demonstrate 
the expressiveness of our data shape abstract 
domain on more examples besides the program 
shown in Figure~\ref{fig:gpa}.

\subsubsection{Magic Trick.} Let us consider the 
following \python program fragment:
\begin{lstlisting}[language=Python]
T = int(input())    
for x in range(T):
    l1 = int (input())
    for i in range(1,l1):
        input()
    L1 = list(map(int,input().split()))
    for i in range(l1+1,5):
        input()
    l2 = int (input())
    for i in range(1,l2):
            input()
    L2 = list(map(int,input().split()))
    for i in range(l2+1,5):
        input()
\end{lstlisting}
(from a solution to the \emph{Magic Trick} problem 
of the Google Code Jam 2014 programming 
competition\footnote{\url{https://codingcompetitions.withgoogle.com/codejam/archive/2014}}).
We instantiate our data shape domain $\adom$ 
with 
the 
type constraining domain $\kdom_\tdom$ and the 
interval constraining domain $\kdom_\mathbb{I}$. 
In this case, our data shape analysis with 
$\adom(\kdom_\tdom, \kdom_\mathbb{I})$, 
determines that correct data files for the program 
have the following shape: 
\begin{equation*}
	\begin{array}{rcrclccc}
		&&&&& \scriptstyle{\textcolor{gray}{1}} & 
		\scriptstyle{\textcolor{gray}{2}} & 
		\scriptstyle{\textcolor{gray}{\dots}} \\
		&&&& \scriptstyle{\textcolor{gray}{1}} & 
		\tuple{\textsc{int}}{[0, \inf]} &  & \\
		\multirow{9}{*}{$d^1_1$} 
		&\multirow{9}{*}{$\begin{cases}
			& \\
			& \\
			& \\
			& \\
			& \\
			& \\
			& \\
			& \\
			&
			\end{cases}$}
		&&& 
		\scriptstyle{\textcolor{gray}{2}} & 
		\tuple{\textsc{int}}{[1, 4]} & 
		& \\
		&& \multirow{3}{*}{$4$} &
		\multirow{3}{*}{$\begin{cases}
			& \\
			& \\
			&
			\end{cases}$}
		&
		\scriptstyle{\textcolor{gray}{3}} & 
		\tuple{\textsc{string}}{[-\inf, \inf]} & 
		\tuple{\textsc{string}}{[-\inf, \inf]} & 
		\dots \\
		&&&& \scriptstyle{\textcolor{gray}{\vdots}} & 
		\dots & 
		\dots & \\
		&&&& \scriptstyle{\textcolor{gray}{6}} & 
		\tuple{\textsc{string}}{[-\inf, \inf]} & 
		\tuple{\textsc{string}}{[-\inf, \inf]} & 
		\dots \\
		&&&&
		\scriptstyle{\textcolor{gray}{7}} & 
		\tuple{\textsc{int}}{[1, 4]} & 
		& \\
		&& \multirow{3}{*}{$4$} &
		\multirow{3}{*}{$\begin{cases}
			& \\
			& \\
			&
			\end{cases}$}
		& \scriptstyle{\textcolor{gray}{8}} & 
		\tuple{\textsc{string}}{[-\inf, \inf]} & 
		\tuple{\textsc{string}}{[-\inf, \inf]} & 
		\dots \\
		&&&& \scriptstyle{\textcolor{gray}{\vdots}} & 
		\dots & 
		\dots & \\
		&&&& \scriptstyle{\textcolor{gray}{11}} & 
		\tuple{\textsc{string}}{[-\inf, \inf]} & 
		\tuple{\textsc{string}}{[-\inf, \inf]} & 
		\dots \\
		&&&& \scriptstyle{\textcolor{gray}{\vdots}} & 
		\dots & 
		\dots & \dots
	\end{array}
\end{equation*}
where $d^1_1$ denotes the first (i.e., and in fact the 
only) data field $1$ of the first 
data 
record in the data file. In particular, we know that 
$1 \leq, d^1_2 \leq 4$ (resp. $1 \leq d^1_7 
\leq 4$) from the for loops at lines 
4-5 and 7-8 (resp. at lines 10-11 and 
13-14). \qee

\subsubsection{Bird Watching.} Let us consider 
now 
the following \python program fragment:
\begin{lstlisting}[language=Python]
N, M, S = map(int, input().split())
pre = [[] for _ in range(N)]
for _ in range(M):
    f, t = map(int, input().split())
    pre[t].append(f)
for n in pre[S]:
    ...
\end{lstlisting}
(from a solution to the \emph{Bird Watching} 
problem of the 
\textsc{SWERC 
	2019-2020} 
programming 
competition\footnote{\url{https://swerc.eu/2019/}}).
We instantiate $\adom$ with the type constraining 
domain $\kdom_\tdom$ and the octagon 
constraining domain $\kdom_\mathbb{O}$. A 
forward numerical
pre-analysis with the octagon domain 
$\mathbb{O}$ determines, in particular, that $0 
\leq 
\mathsf{len}(pre) \leq N - 1$ (cf. line 2).
Thus, our backward data shape analysis with 
$\adom(\kdom_\tdom, \kdom_\mathbb{O})$ 
determines that correct data files for the program 
have the following shape: 
\begin{equation*}
	\begin{array}{rclccc}
		&&& \scriptstyle{\textcolor{gray}{1}} & 
		\scriptstyle{\textcolor{gray}{2}} & 
		\scriptstyle{\textcolor{gray}{3}} \\
		&& \scriptstyle{\textcolor{gray}{1}} & 
		\tuple{\textsc{int}}{0 
			\leq d^1_1 } & \tuple{\textsc{int}}{0 
			\leq d^2_1 } & \tuple{\textsc{int}}{0 
			\leq d^3_1 \leq d^1_1 - 1 } \\
		\multirow{3}{*}{$d^2_1$} 
		&\multirow{3}{*}{$\begin{cases}
			& \\
			& \\
			&
			\end{cases}$}
		& 
		\scriptstyle{\textcolor{gray}{2}} & 
		\tuple{\textsc{int}}{\text{true}} & 
		\tuple{\textsc{int}}{0 
			\leq d^2_2 \leq d^1_1 } & \\
		&& \scriptstyle{\textcolor{gray}{3}} & 
		\tuple{\textsc{int}}{\text{true}} & 
		\tuple{\textsc{int}}{0 
			\leq d^2_3 \leq d^1_1 } & \\
		&& \scriptstyle{\textcolor{gray}{\vdots}} & 
		\dots & 
		\dots &
	\end{array}
\end{equation*}
where $d^j_i$ denotes the data field $j$ of the 
data 
record $i$.
In particular, we know that $0 
\leq d^3_1 \leq d^1_1 - 1$ from the list access at 
line 6 and, similarly, $0 
\leq d^2_i \leq d^1_1$, for $2 \leq i$, from the list 
access at line 5. \qee

\subsubsection{Adult Census Data.}
Let us consider the following fragment of a 
pre-processing \python function for the Adult 
Census
dataset\footnote{\url{https://archive.ics.uci.edu/ml/datasets/adult}}:
\begin{lstlisting}[language=Python]
def pre_process_data (data):
    new_data = []
    for i in range(len(list(data))):
        person_new = [] 
        
        person_new.append(data[i][0])
        
        w = data[i][1]
        if w == "Private":
           person_new.append(w)
        elif w == "Self-emp-not-inc" or w == "Self-emp-inc":
             person_new.append("Self-Employed")
        elif w == "Federal-gov" or w == "Local-gov" or w == "State-gov":
             person_new.append("Government")
        elif w == "Without-pay" or w == "Never-worked":
             person_new.append("Other")
        else: raise Exception("Workclass not matched: ", w, i)         
        
        ...
        
        new_data.append(person_new)
        
    return new_data
\end{lstlisting}
(taken from \cite{Urban19})
where the function argument 
\lstinline[language=Python]{data} has 
been 
loaded from a CSV file. 
Our 
backward shape analysis instantiated with the  
string set constraining domain 
$\kdom_\wdom$
determines that 
correct 
CSV files have the following shape:
\begin{equation*}
	\begin{array}{lccc}
		& \hspace{3em} 
		\scriptstyle{\textcolor{gray}{1}} & 
		\hspace{3em}
		\scriptstyle{\textcolor{gray}{2}} & 
		\hspace{3em} 
		\scriptstyle{\textcolor{gray}{\dots}} \\
		\scriptstyle{\textcolor{gray}{1}} & 
		\hspace{3em} \top_\wdom & \hspace{3em} W 
		& 
		\hspace{3em} \dots \\
		\scriptstyle{\textcolor{gray}{2}} & 
		\hspace{3em} \top_\wdom & 
		\hspace{3em} W & \hspace{3em} \dots \\
		\scriptstyle{\textcolor{gray}{3}} & 
		\hspace{3em} \top_\wdom & 
		\hspace{3em} W & \hspace{3em} \dots \\
		\scriptstyle{\textcolor{gray}{\vdots}} & 
		\hspace{3em} \dots & 
		\hspace{3em} \dots & \hspace{3em} \dots
	\end{array}
\end{equation*}
where $W$ is the set of strings $\{$
'Private', 'Self-emp-not-inc', 
'Self-emp-inc', 'Federal-gov', 'Local-gov', 
'State-gov', 'Without-pay', 'Never-worked'
$\}$. \qee

\section{Related Work}\label{sec:related}

Learning the input format of a given program is not 
a new
research area but it has recently seen increased 
interest, 
especially in the contest of grammar-based 
automated test 
generation and 
fuzzing applications 
\cite[etc.]{Godefroid08,Holler12,Majumdar07}.

Many of the approaches in the literature are 
\emph{black-box}, e.g., \textsc{glade} 
\cite{Bastani17} and \textsc{Learn\&Fuzz} 
\cite{Godefroid17}. These generally generate input 
grammars or grammar-like structures that are 
strictly meant as intermediate representation to be 
fed to a test generation engine and are not meant to 
be readable by a human. On the other hand, the 
result of our analysis is human-readable and can be 
used for other purposes than test generation, e.g., 
code specification and data cleaning. Moreover, 
these approaches have to rely on samples 
of valid inputs, while our approach only needs the 
program to be analyzed.

Another sample-free approach is \textsc{autogram} 
\cite{Hoschele16}, which uses dynamic data flow
analysis to generate readable and usable grammars. 
One disadvantage of this approach is that it will 
skip parts of the input if these are not stored in 
some program variables (e.g. if a program scans 
over a comment). On the contrary, in such a case, 
our approach will 
not know any value information about the skipped 
data but will still know that this data should be 
present in the data file (see the \emph{Magic Trick} 
example in Section~\ref{sec:implementation} for 
instance).

To the best of our knowledge ours is the first 
approach that uses static analysis 
to infer the input format of a given program. 
Moreover, contrary to the above grammar synthesis 
approaches, our approach infers \emph{semantic} 
(and not just syntactic)
information on the input data of a program.
Closest to ours, is the work of Cheng and Rival 
\cite{Cheng15} on the static analysis of 
spreadsheet 
applications. They however only focused on 
type-related properties.

Finally, the main difference compared to the 
inference of necessary preconditions proposed by 
Cousot et al. \cite{Cousot13} or the (bi-)abduction 
\cite{Calcagno11} used in tools like \textsc{Infer} 
\cite{Calcagno15} is that our analysis can also deal 
with inputs read at any point during the program 
(thus notably also inside loops whose execution 
may depend on other inputs --- this is where the 
need for the stack comes from, cf. 
Section~\ref{sec:stack}). 

\section{Conclusion and Future 
Work}\label{sec:conclusion}

In this paper, we have proposed a parametric static 
shape 
analysis framework based on abstract interpretation 
for inferring semantics properties of
input data of data-processing programs. 
Specifically, our 
analysis automatically infers necessary conditions 
on the 
structure and values of the input data for the 
data-processing program to run successfully and 
correctly.

It remains for future work to explore possible 
applications of the result our analysis. In particular, 
we are interested in developing better
grammar-based testing approaches. We are also 
interested in developing tools for assisting and 
guiding or even 
automating the checking and cleaning of data. 


\bibliographystyle{abbrv}
\bibliography{bibliography}

\end{document}